\documentstyle[12pt]{article}
\begin{document}

\baselineskip 6mm
\renewcommand{\thefootnote}{\fnsymbol{footnote}}

\newcommand{\nc}{\newcommand}
\newcommand{\rnc}{\renewcommand}

%\headheight=0truein
%\headsep=0truein
%\topmargin=0truein
%\oddsidemargin=0truein
%\evensidemargin=0truein
%\textheight=9truein
%\textwidth=6.5truein

\rnc{\baselinestretch}{1.24}    % 1.5 spacing btwn text lines
\setlength{\jot}{6pt}       % spacing btwn the rows of an eqnarray
\rnc{\arraystretch}{1.24}   % spacing btwn the rows of a non-eqn array

%%%%%%%%%%%%%%%%%%%%%% Equation Numbering %%%%%%%%%%%%%%%%%%%%%%%
%\makeatletter
%\rnc{\theequation}{\thesection.\arabic{equation}}
%\@addtoreset{equation}{section}
%\makeatother

%%%%%%%%%%%%%%%%%%%%%%%%%%%%%%%%%%%%%%%%%%%%%%%%%%%%%%%%%%%%%%%%%
%                                                               %
%                NEW COMMANDS AND MACROS                        %
%                                                               %
%%%%%%%%%%%%%%%%%%%%%%%%%%%%%%%%%%%%%%%%%%%%%%%%%%%%%%%%%%%%%%%%%

%%%%% Simplify some frequently used LaTeX commands %%%%%

\nc{\be}{\begin{equation}}

\nc{\ee}{\end{equation}}

\nc{\bea}{\begin{eqnarray}}

\nc{\eea}{\end{eqnarray}}

\nc{\ben}{\begin{eqnarray*}}

\nc{\een}{\end{eqnarray*}}

\nc{\xx}{\nonumber\\}

\nc{\ct}{\cite}

\nc{\la}{\label}

\nc{\eq}[1]{(\ref{#1})}

\nc{\newcaption}[1]{\centerline{\parbox{6in}{\caption{#1}}}}

\nc{\fig}[3]{

\begin{figure}
\centerline{\epsfxsize=#1\epsfbox{#2.eps}}
\newcaption{#3. \label{#2}}
\end{figure}
}

%%% Double line letters %%%

\def\IR{{\hbox{{\rm I}\kern-.2em\hbox{\rm R}}}}
\def\IB{{\hbox{{\rm I}\kern-.2em\hbox{\rm B}}}}
\def\IN{{\hbox{{\rm I}\kern-.2em\hbox{\rm N}}}}
\def\IC{\,\,{\hbox{{\rm I}\kern-.59em\hbox{\bf C}}}}
\def\IZ{{\hbox{{\rm Z}\kern-.4em\hbox{\rm Z}}}}
\def\IP{{\hbox{{\rm I}\kern-.2em\hbox{\rm P}}}}
\def\IH{{\hbox{{\rm I}\kern-.4em\hbox{\rm H}}}}
\def\ID{{\hbox{{\rm I}\kern-.2em\hbox{\rm D}}}}

%%%%% Roman pont in math

\def\Tr{{\rm Tr}\,}
\def\det{{\rm det}}

%%%%% Special Letters

\def\vare{\varepsilon}
\def\barz{\bar{z}}
\def\barw{\bar{w}}

\begin{titlepage}
%---------------- preprint number ---------------
\hfill\parbox{5cm}
{DAMTP-2004-53 \\ IP/BBSR/2004-11 \\ hep-th/0405261}\\
\vspace{25mm}
\begin{center}
%------------------------ title ------------------------
{\Large {\bf Gravitational wave solutions in string and\\
M-theory AdS Backgrounds}  }

\vspace{15mm}
%---------------- authors and addresses ----------------
Alok Kumar$^{1}$\footnote{kumar@iopb.res.in} and Hari K.
Kunduri$^{2}$\footnote{H.K.Kunduri@damtp.cam.ac.uk}
\\[3mm]
{\sl 1. Institute of Physics, Bhubaneswar 751 005, INDIA} \\
{\sl 2. DAMTP, University of Cambridge, Cambridge, United Kingdom}

\end{center}

\thispagestyle{empty}

\vskip2cm

%----------------------- abstract ----------------------

\centerline{\bf ABSTRACT}
\vskip 4mm
\noindent
In this paper, we present several gravitational wave solutions
in $AdS_5\times S^5$ string backgrounds, as well as in
$AdS_7\times S^4$ and  $AdS_4\times S^7$ backgrounds in
M-theory, generalizing the results of hep-th/0403253 by
one of the authors. In each case, we present the general form of such
solutions and give explicit examples, preserving certain
amount of supersymmetry, by taking
limits on  known BPS D3  and M2, M5-brane solutions in pp-wave backgrounds.
A key feature of our examples is the possibility of a wider
variety of wave profiles, than in pure gravity
and string/M-theory examples known earlier,
coming from the presence of various p-form field strengths
appearing in the gravitational wave structure.
\\

\vspace{2cm}

\today

\end{titlepage}

\section{Introduction}

Gravitational waves have long been a subject of research
\cite{brinkmann,stefani}. Such solutions in general relativity,
also known as pp-waves, are mostly discussed in the context of
asymptotically flat Minkowski spaces.
On the other hand, exact solutions  of Einstein equations
representing gravitational waves in non-asymptotically flat
backgrounds have also been analyzed
\cite{carmelli,podolsky1,podolsky2} over a long period of time.
Classical solutions representing gravitational waves provide a
geometrical framework to understand gravitational radiation
and may thus have astrophysical implications as well.

In this paper, gravitational wave solutions are obtained
in various anti-de Sitter backgrounds in string
and M-theory~\cite{cvetic,brecher,patri,kumar}.
In this context, we
extend the results of a previous paper by one of the
authors in \cite{kumar}, where gravitational waves in
$AdS_3\times S^3\times R^4$ were discussed,
and write down several general solutions in $AdS_5\times S^5$
string backgrounds, as well as $AdS_4\times S^7$ and
$AdS_7\times S^4$ backgrounds in
M-theory. In each of these cases, we also give explicit examples
by applying scaling limits~\cite{maldacena,khastgir} (also
identified as the near-horizon geometry)
on known D3 and M2, M5-brane solutions
\cite{KNS-etc,alishahiha,biswas,mas} in pp-wave backgrounds.
In particular, two explicit examples of gravitational waves in
$AdS_5\times S^5$ are obtained by taking limits on known
(supersymmetric) D3-branes given in \cite{alishahiha,biswas}.
In the first of these cases, in section (2.1), the gravitational wave
profile ($H$) turns out to depend only on $AdS_5$ coordinates, implying
that only such graviton polarizations may exist for the
gravitational wave, a fact normally seen by transforming to the
`Rosen' coordinates. In the second $AdS_5\times S^5$ example in
section-(2.2), however, we have $H$ depending on $S^5$ coordinates
as well. Examples for the $AdS_7\times S^4$ gravitational wave in
section-3  and $AdS_4\times S^7$ gravitational wave in section-4
are respectively obtained from $M5$ brane solutions in
\cite{singh,alishahiha,mas} and $M2$ brane solution in \cite{mas}.
As will be noticed, guided by the general form of the
$AdS$ examples constructed  from the D-branes and M-branes, we are able to
write a general class of solutions (with certain constraints)
in each case. Supersymmetry is, however, discussed only for
the special cases.

Earlier work on string theory in pp-wave backgrounds
\cite{amati-horowitz} are summarized in \cite{tseytlin}. Recent
developments~\cite{blau,metsaev,BMN}, including 
`Penrose limits'~\cite{penrose} and applications to
four dimensional gauge theories\cite{BMN}, are reviewed in
\cite{sheikh}. Some other aspects of D-branes in pp-wave
backgrounds are discussed in \cite{dabholkar}. A crucial feature
of our solutions, compared to those in \cite{cvetic,brecher}, is
the presence of p-form field strengths affecting the structure of
the gravitational waves. In other words, the gravitational wave
equations in earlier examples~\cite{cvetic,brecher,patri} are identical
to those appearing in a higher dimensional $AdS\times S$ pure gravity
theory with a cosmological constant. The role of string or
M-theory in those solutions is only to provide a consistent
background configuration without affecting the wave nature. In our
examples, extending the results of \cite{kumar}, however, the
p-form fields appear in the wave equations and thus provide room
for a wider variety of solutions.

\section{$AdS_5\times S^5$ gravitational waves}

\subsection{$AdS_5$ Wave profile}

We begin our discussion by considering first an example of
gravitational wave in the $AdS_5\times S^5$ background in string
theory. The general form of the metric used in this case has a
form: 
\be 
ds^2 = q \left\{ {du^2\over u^2} + {1\over u^2} (2 dx^+ dx^-
       + H(u, x^+, x^i, x^a) {dx^+}^2 + \sum_{i=1}^2 {dx^i}^2) +
       d\Omega_5^2 \right\},
\label{metric} 
\ee 
where the parameter $q$ gives the radius of
curvature for the $AdS_5$ and $S^5$ spaces. In our examples,
coordinates $x^{+}$, $x^{-}$, $u$ and $x^{i}$'s run over $AdS$
directions whereas $x^a$'s denote the directions along $S^n$. The
metric in equation (\ref{metric}) is a generalization of the one
for the $AdS_4$ case in general
relativity~\cite{podolsky1,podolsky2} to $AdS_5\times S^5$. We will
use a similar form of the metric for M-theory $AdS_7$ and $AdS_4$
examples as well.

For the general solution above, we have assumed that the
wave profile $H$ in the above metric may depend on all the
AdS as well as $S^5$ coordinates. However, specific coordinate dependence
of $H$ also gives the graviton polarizations that are turned on.
In the present example, we also
take an NS-NS 3-form flux of the form:
\be
  H^{(3)} = dx^+ \wedge
  \left( ({- 2\mu q A_{i} (x^i, x^a, x^+) \over u^2}) du
            \wedge d x^i + ({2\mu q A_{i a} (x^i, x^a, x^+) \over u})
             dx^i \wedge dx^a \right),
\label{nsns3-form}
\ee
and an R-R 3-form flux:
\be
  F^{(3)} = dx^+ \wedge
  \left( ({- 2\mu q B_{i} (x^i, x^a, x^+) \over u^2}) du
            \wedge d x^i + ({2\mu q B_{i a} (x^i, x^a, x^+) \over u})
             dx^i \wedge dx^a \right),
\label{rr3-form}
\ee
with $\mu$ being a parameter characterizing the gravitational wave.
By setting $\mu =0$ one goes to the $AdS_5\times S^5$ background solution.

One also has the usual $R-R$ 5-form flux necessary for constructing
the background $AdS_5\times S^5$ solution in type IIB string theory:
\be
F^{(5)} =  - {4 q^2 \over u^5}dx^+ \wedge dx^- \wedge du
           \wedge dx^1 \wedge dx^2
          -  4 q^2 \;\sqrt{g}\; d\theta^1 \wedge d\theta^2
             \wedge d\theta^3 \wedge d\theta^4 \wedge d\theta^5,
\label{rr5-form}
\ee
where $\theta^i$'s are five angular coordinates on $S^5$ and
$g$ is the determinant of the metric ($g_{ab}$)
on this space. Later on,
while presenting the explicit example, we will also give expressions for
the metric $g_{a b}$ in terms of angles $\theta_1,...,\theta_5$.
 
Ricci curvature tensors for the $AdS_5$ components for the 
above metric have the form:
\bea
   R_{++} = - {H_{, uu}\over 2} + {3\over 2}{H_{, u}\over u}
            - {4H\over u^2} - {1\over 2} H_{,i}^i -
             {1\over 2u^2\sqrt{g}}\partial_a
       (\sqrt{g}\partial^a H), \cr
&\cr
   R_{+-} = R_{uu} = - {4\over u^2}, \;\;
   R_{ij} = - {4\delta_{ij}\over u^2},\; 
\label{ricci-ads5}
\eea
and Ricci tensor for the $S^5$ components satisfy:
\be
    R_{a b} = 4 g_{a b}.
\label{ricci-s5}
\ee

Now, before starting to solve the full type IIB equations of motion
(see for example , \cite{oz}), we make a few comments regarding the form
of the strss energy tensor for the ansatz presented above. In 
particular, we note that although the three-from flux 
in equations (\ref{nsns3-form}) and (\ref{rr3-form}) are  
inhomogeneous in the sphere direction due to the $x^a$ dependence in 
$A_i, B_i$ and $A_{ia}, B_{ia}$, but they are {\em{null}} fluxes. 
Also, from our above metric, one easily deduces that $g^{+\mu} = 0$ unless 
$\mu = -$. Thus, when one contracts over the flux indices, for example, 
in a term such as $F_{\mu \nu \delta}F^{\mu \nu \delta}$,  the 
result is zero. This is because none of these fluxes have a 
$dx^{-}$ leg. Since the stress energy tensor coming from the 
various 3-form fluxes generically has a form:
\begin{equation}
T_{ab} = \alpha F_{acd}F_{b}^{\phantom{b}cd} + \beta
F_{efg}F^{efg}g_{ab},
\end{equation} where $\alpha$ and $\beta$ are constants, one
immediately reads off that the second term will always be zero, for
all $(a,b)$ for any flux that has a $dx^{+}$ leg.  Further, by the
exact same reasoning as above, the first term must vanish unless 
$(a,b) = (++)$. One deduces quickly that only the $T_{++}$ components 
are modified by the introudciton of our null fluxes, relative to the 
original ($H=0$) solution.

Furthermore, note the introduction of the wave term only affects the 
Ricci tensor of the background $AdS_{5} \times S^{5}$ by introducing a 
non-zero $R_{++}$, as we have written explicitly above in 
equations (\ref{ricci-ads5}) and (\ref{ricci-s5}). It is for this 
reason that we have chosen our null fluxes in the form~(\ref{nsns3-form}) 
and~(\ref{rr3-form}); their contriubitoin to $T_{++}$ provides a source 
for $R_{++}$ via the Einstein equations. For a thorough discussion of  
these points, see the discussion preceeding equation (2.10) in ~\cite{BDGO}.

We now proceed further and analyze the field equations and their 
solutions. The type IIB string theory equations of motion \cite{oz} 
reduce to the following conditions for our ansatz (in eqns. 
(\ref{metric}), (\ref{nsns3-form}), (\ref{rr3-form}) and
(\ref{rr5-form})):
\be
    - {H_{, uu}\over 2} + {3\over 2} {H_{, u}\over u}
    - {1\over 2} H_{,i}^i - {1\over 2u^2\sqrt{g}}\partial_a
       (\sqrt{g}\partial^a H)
         =  {2\mu^2} \left( \sum_i [A_i^2 + B_i^2] +
             \sum_a [A_{ia}A_i^a + B_{ia} B_i^a]  \right),
\label{h-condition}
\ee
\be
   \partial_i A_i\; =\; \partial_i A_{ia}\; = \;\partial_i B_i
    = \partial_i B_{i a} = 0,
\label{AB-condition1}
\ee
\bea
\nabla^a A_{ia} - A_i + 4\epsilon_{ij} B_j =0,\cr
\nabla^a B_{ia} - B_i - 4\epsilon_{ij} A_j =0.
\label{AB-condition2}
\eea
The Bianchi identity on $H^{(3)}$ and $F^{(3)}$ imply the following conditions:
\be
    \partial_a A_i = - A_{i a},\;\;
    \partial_a B_i = - B_{i a}.
\label{bianchi}
\ee

Now, several explicit solutions for the above set of
conditions can be obtained. First, we write down the
type of solution already
known in the literature. They correspond to the choice: $A_i = B_i
= A_{ia} = B_{ia} = 0$ and originate from certain
`brane waves'\cite{cvetic,brecher,patri}. Then, if one chooses
$H$ to be dependent only on coordinate `$u$' and $x^+$, we get
a solution:
\be
H_0 = f_o(x^+) u^4 + f_1(x^+),
\label{solutionH0}
\ee
with $f_{0}$, $f_1$ being arbitrary
functions of $x^+$. One can also add to $H_0$
harmonic functions $H_1$ and $H_2$ satisfying $H,i^i = 0$ and
$\nabla^a\partial_a H = 0$ respectively. Several other solutions can be
obtained by taking products of functions of the type
$H_0$, $H_1$ and $H_2$.

We now discuss the new solutions which emerge due to
the presence of nontrivial NS-NS and R-R 3-form field strengths
in the gravitational wave equation. To write down a solution
explicitly, we make a choice for the metric on $S^5$ as:
\be
   d\Omega_5^2 = d\theta^2 + {\rm{sin}}^2\theta\; d\phi^2 +
                 \rm{cos}^2\theta\; d\alpha^2 + \rm{cos}^2\theta\;
                 \rm{cos}^2\alpha\; d\beta^2 + \rm{cos}^2\theta\;
                 {\rm{sin}}^2\alpha\; d\gamma^2.
\label{deomega5}
\ee
These angular coordinates are related to the six dimensional
Cartesian coordinates as:
\bea
&&y_1 = r\; {\rm{sin}}\theta\; {\rm{cos}}\phi, \;\;
y_2 = r\; {\rm{sin}}\theta\; {\rm{sin}}\phi, \;\;
y_3 = r\; {\rm{cos}}\theta\; {\rm{cos}}\alpha\; {\rm{cos}}\beta,\cr
&  \cr
&&y_4 = r\; {\rm{cos}}\theta\; {\rm{cos}}\alpha\; {\rm{sin}}\beta,\;\;
y_5 = r\; {\rm{cos}}\theta\; {\rm{sin}}\alpha\; {\rm{cos}}\gamma,\;\;
y_6 = r\; {\rm{cos}}\theta\; {\rm{sin}}\alpha\; {\rm{sin}}\gamma.
\label{coordinates}
\eea
The solution for $A_i$, $A_{i a}$, $B_i$, $B_{ia}$ in equations
(\ref{metric}), (\ref{nsns3-form}), (\ref{rr3-form}) and
(\ref{rr5-form}) then are:
\bea
&&A_1 = {\rm{sin}}\theta\; {\rm{cos}}\phi,\;\;
A_2 = {\rm{sin}}\theta\; {\rm{sin}}\phi,\;\cr
& \cr
&&A_{1\theta} = - {\rm{cos}}\theta\; {\rm{cos}}\phi,\;\;
A_{1\phi} = {\rm{sin}}\theta\; sin\phi,\cr
& \cr
&&A_{2\theta} = - {\rm{cos}}\theta\; sin\phi,\;\;
A_{2\phi} = - {\rm{sin}}\theta\; cos\phi,
\label{Ai-Aia}
\eea
and
\bea
&&B_1 = {\rm{sin}}\theta\; {\rm{sin}}\phi,\;\;
B_2 = - {\rm{sin}}\theta\; {\rm{cos}}\phi,\;\cr
& \cr
&&B_{1\theta} = - {\rm{cos}}\theta\; {\rm{sin}}\phi,\;\;
B_{1\phi} = - {\rm{sin}}\theta\; {\rm{cos}}\phi,\cr
& \cr
&&B_{2\theta} =  {\rm{cos}}\theta\; {\rm{cos}}\phi,\;\;
B_{2\phi} =  - {\rm{sin}}\theta\; {\rm{sin}}\phi.
\label{Bi-Bia}
\eea
Also,the 5-form field strength can also be written in terms of
angular variables $\theta$, $\phi$, $\alpha$, $\beta$ and $\gamma$
for the above parameterization 
by using $\sqrt{g} = {\rm{sin}}\theta\; {\rm{cos}}^3\theta\;
{\rm{sin}}\alpha\; {\rm{cos}}\alpha$.

For the choice of the metric in $(\ref{deomega5})$, very few nonzero
components of the Christoffel connection exist, namely
$\Gamma^{\theta}_{\phi\phi}$, $\Gamma^{\theta}_{\alpha\alpha}$,
$\Gamma^{\theta}_{\beta\beta}$. $\Gamma^{\theta}_{\gamma\gamma}$, 
$\Gamma^{\alpha}_{\beta\beta}$, $\Gamma^{\alpha}_{\gamma\gamma}$ 
and
$\Gamma^{\phi}_{\theta\phi}$, $\Gamma^{\alpha}_{\theta\alpha}$,
$\Gamma^{\beta}_{\theta\beta}$, $\Gamma^{\gamma}_{\theta\gamma}$,
$\Gamma^{\beta}_{\alpha\beta}$, $\Gamma^{\gamma}_{\alpha\gamma}$.
We skip their detailed form here. It can now be directly verified that
$A_i$, $A_{i a}$, $B_i$, $B_{ia}$ given in equations
(\ref{Ai-Aia}) and (\ref{Bi-Bia}) satisfy equations (\ref{AB-condition1}),
(\ref{AB-condition2}) and (\ref{bianchi}). Moreover, we also have:
\be
\sum_i [A_i^2 + B_i^2 +  \sum_a (A_{ia}A_i^a + B_{ia} B_i^a) ] = 4,
\label{Ai-square}
\ee
leading to the following equation for the wave profile $H$
(by using equation (\ref{h-condition})):
\be
- {H_{, uu}\over 2} + {3\over 2} {H_{, u}\over u}
    - {1\over 2} H_{,i}^i - {1\over 2u^2\sqrt{g}}\partial_a
       (\sqrt{g}\partial^a H) = 8\mu^2.
\label{H-equation1}
\ee
The condition (\ref{H-equation1}) again has several solutions.
When  $H$ is a function of $u$ and $x^+$ only, one has
\be
    H_1 = 4\mu^2 u^2 + f_0(x^+),
\label{H0-solution}
\ee
with $f_0(x^+)$ being an arbitrary  function of $x^+$ only.

One can, instead, take $H$ to be a function of coordinates
$x^i$ and $x^+$ only, leading to
\be
   H_2 = - 4\mu^2 \sum_{i=1}^2 {x^i}^2 +
\sum_{i=1}^2  f_i(x^+)x^i + g (x^+),
\label{H1-solution}
\ee
with $f_1, f_2$ and $g$ being functions of $x^+$ only. One can also take
linear combinations of solutions of the types in equations
(\ref{H0-solution}) and (\ref{H1-solution}) with coefficients
$a_0$ and $a_1$, such that $a_0 + a_1 = 1$.

We can further generalize the solutions by multiplying
$A_i, A_{ia}$, $B_i, B_{ia}$ in equations (\ref{Ai-Aia}) and
(\ref{Bi-Bia}) by an arbitrary function of $x^+$: $F(x^+)$,
while simultaneously multiplying $H$ by another function
$G(x^+)$ such that $G = F^2$.

We now show that for a special case in equation
(\ref{H1-solution}), namely
\be
H \equiv \hat{H} = - 4\mu^2 ({x^1}^2 + {x^2}^2),
\label{H1'}
\ee
(obtained by setting $f_{1, 2}$ and $g$ to zero in $H_2$),
the gravitational wave solution follows from a (singular)
scaling  limit  of a supersymmetric $D3$ brane solution in a pp-wave
background\cite{alishahiha}. Similar limits have been applied 
to obtain other examples of gravitational  wave solutions
in Ads backgrounds previously~\cite{cvetic,brecher,patri,kumar}.

The (localized) $D3$ brane solution in a pp-wave background,
giving the above gravitational wave solution,
in a scaling limit : $r\rightarrow 0$,
with a wave profile $\hat{H}$, is
written  as~\cite{alishahiha}:
\bea
ds^2&=&f^{-{1\over 2}}\left(2dy^+dy^-
-4\mu^2[{\tilde{y}_2}^2+ {\tilde{y}_4}^2](dy^+)^2+ d{\tilde{y}^2}_2
+d{\tilde{y}_4}^2\right)
+f^{{1\over 2}}\left(dr^2+r^2d\Omega_5^2\right),\cr &&\cr
F_{+32} &=&  F_{+41}=2\mu,
\;\;\;\;\; B_{+1}=2\mu \tilde{y}_2,\;\;\;\;\;\;
B_{+3}=2\mu \tilde{y}_4\cr &&\cr
F_{mnlpq}&=&\epsilon_{mnlpqs}\partial_{s}f,\;\;\;\;\;\;
f=1+{q^2\over r^4},\;\;\;q^2 = c_3Ng_sl_s^4.
\label{D3soln.}
\eea
This is a D3 brane solution in a pp-wave background specified by the
parameter $\mu$. When one sets the $D3$ brane charge to zero, one gets
the background metric of the pp-wave, with NS-NS and R-R 3-form flux
having components, $H_{+12}, H_{+34}$ and $F_{+23}, F_{+14}$
respectively. In the above D3 brane solution, directions
$\tilde{y}^2, \tilde{y}^4$ are longitudinal coordinates of the 
brane, whereas
$\tilde{y}^1, \tilde{y}^3, \tilde{y}^5,...,\tilde{y}^8$ are
transverse to the brane.

To obtain the gravitational wave metric in $AdS_5\times S^5$
background as given in equation (\ref{metric}), from the D3 brane solution
above, we take $r \rightarrow 0$ limit in the Green function ($f$) in
equation (\ref{D3soln.}). One then obtains the metric in equation
(\ref{metric}) with the profile $H = \hat{H}$ given as in equation
(\ref{H1'}), when one also identifies:
\be
   x^1 \equiv \tilde{y}^2,\;\;x^2 \equiv \tilde{y}^4,\;
   y^1 \equiv \tilde{y}^1,\;\;y^2 \equiv \tilde{y}^3,\;\;
   y^{3..6} \equiv \tilde{y}^{5..8},
\label{redefn}
\ee
and defines $r = {q\over u}$. 
For the $D3$ brane solution in equation
(\ref{D3soln.}), the radial coordinate along transverse direction is:
$r^2 = {\tilde{y^1}}^2 + {\tilde{y^3}}^2 +
\sum_{i=5}^8 {\tilde{y^i}}^2 $.
In the new variables that we are using, one gets:
$r^2 = \sum_{i=1}^6 {y^i}^2$. NS-NS and R-R 3-form field strengths
in equations (\ref{nsns3-form}) and (\ref{rr3-form}) (with $A$'s
and $B$'s as in equations (\ref{Ai-Aia}) and (\ref{Bi-Bia})) also arise
from the ones in equation (\ref{D3soln.}) by redefining the coordinates as
in equation (\ref{redefn}) and making use of the angular coordinates in
equation (\ref{coordinates}). By such change of variables, one obtains
the expressions in equations 
(\ref{nsns3-form}) and (\ref{rr3-form}) with
solutions for $A_i, A_{ia}, B_i, B_{ia}$ as in equations
(\ref{Ai-Aia}), (\ref{Bi-Bia}).

We have therefore obtained a general class of
$AdS_5\times S^5$  gravitational
wave solution in type IIB string theory and also presented an explicit
example characterized by functions $A_i, A_{i a}$, $B_i, B_{ia}$
in equations (\ref{Ai-Aia}), (\ref{Bi-Bia}) and the wave profile
$\hat{H}$ in equation (\ref{H1'}). We have also shown how our 
gravitational wave arises from a $D3$-brane in a pp-wave
background. This connection with $D3$-brane has been
presented for the wave 
profile $\hat{H}$. It will be interesting to see if the 
wave profile $H_1$ can also arise from a D-brane in a similar way.

We now discuss the supersymmetry property of the solutions described above.
To show that the above gravitational wave, with 
$H$ (= $\hat{H}$) as in equation (\ref{H1'}), is supersymmetric, 
one also notes that
the original $D3$ brane given in equation (\ref{D3soln.})
preserves a certain amount of
supersymmetry as well.\footnote{In solution
(\ref{D3soln.}) above we have corrected a minus sign in one of the
R-R 3-form component in \cite{alishahiha}.}
Therefore the limiting solution, appearing as gravitational wave
above, is expected to be supersymmetric as well.

To elaborate more, the Killing spinors~\cite{alishahiha} for the 
$D3$ brane solution~\cite{alishahiha} 
are the ones which satisfy the following 
projections: (1) the $D3$ brane supersymmetry condition identical to the one 
in flat space, relating $\epsilon_{\pm}$ with $\epsilon_{\mp}$, 
(2) the usual $\Gamma^{\hat{+}}$ projection of the gravitational wave 
solution implying either a left-moving or a right-moving wave, 
(3) An additional condition due to the presence of NS-NS and R-R 3-form 
flux. For the relevant $D3$ brane discussed above, these conditions
are mentioned in equations (42), (38) and (43) of \cite{alishahiha}. 
The important point to note is that these conditions  are independent of
the function $f$, namely the Green function in the transverse space. 
As a result, the limiting procedure that we described for getting
the $AdS_5\times S^5$ gravitational wave gives the identical 
Killing spinors as the $D3$ brane case. In the present case, 
however, one also expects the presence of additional Killing 
spinors, since as is known for flat $D3$ branes, the supersymmetry 
enhances in the near horizon limit from one half to the maximal supersymmetry.
Finding the exact amount of supersymmetry, by writing down all the 
Killing spinors explicitly is important.
We, however, do not go into them right now
and simply end the section by saying that our gravitational wave
solution is supersymmetric for the wave profile $\hat{H}$, (with 
nontrivial NS-NS and R-R 3-form flux as given in equations (\ref{Ai-Aia})
and (\ref{Bi-Bia})). The case of other wave profile, namely  
$H_1$ is also expected to be supersymmetric, even though the 
$D3$ brane connection is not apparent for this example.

\subsection{Wave profile with $S^5$ dependence}

We now give an example of the gravitational wave where the wave profile
$H$ depends on $S^5$ coordinates as well. An explicit example of this
type originates from a $D3$ brane solution in \cite{biswas}. The
general form of the metric for this solution is same as in
equation (\ref{metric}). The 5-form field is also identical to the one
in equation (\ref{rr5-form}). However, one now has only an NS-NS 
3-form flux of the form:
\be
  H^{(3)} = dx^+ \wedge
  \left( ({- \mu q^2 A_{a} (x^i, x^a, x^+) \over u^3}) du
            \wedge d x^a + ({\mu q^2 A_{a b} (x^i, x^a, x^+) \over u^2})
             dx^a \wedge dx^b \right).
\label{nsns3-form22}
\ee
One can also generate a combination of NS-NS and R-R
3-form field strengths by using
$S$ duality symmetry of the IIB string theory. We, however, do not
go into this aspect here. One also notices that the general form
of the NS-NS 3-form field strength above is identical to the one used in
\cite{kumar} for the $AdS_3\times S^3$ example.

In the present case, the gravitational wave is therefore characterized
by the metric, R-R 5-form flux and NS-NS
3-form flux as in equations (\ref{metric}), (\ref{rr5-form}) and
(\ref{nsns3-form22}) respectively.
The equation for the wave profile $H$ is now given
as:
\be
    - {H_{, uu}\over 2} + {3\over 2} {H_{, u}\over u}
    - {1\over 2} H_{,i}^i - {1\over 2u^2\sqrt{g}}\partial_a
       (\sqrt{g}\partial^a H)
         =  {\mu^2 q^2\over 2u^4} \left( \sum_a A_a A^a +
            {1\over 2} \sum_{a,b} (A_{ab}A^{ab})    \right).
\label{h-condition22}
\ee
One also has additional conditions on quantities $A_a$ and 
$A_{a b}$\cite{kumar}:
\be
\nabla^a A_{a} =0,\;\;
\nabla^b A_{ab} = 4 A_a,
\label{AB-condition22}
\ee
coming from equations of motion. The Bianchi identity
implies\cite{kumar}:
\be
    \partial_{[a} A_{b]} =  A_{a b}\;.
\label{bianchi22}
\ee
Although the condition (\ref{bianchi22}) coming from the
Bianchi identity for this example is
identical to the one for $AdS_3\times S^3\times R^4$ case in \cite{kumar}
due to the identical form of the NS-NS 3-form in the two cases,
one of the equation of motion
in (\ref{AB-condition22}) has a different factor than in \cite{kumar}
due to the presence of
additional coordinates $x^{1,2}$ in $AdS_5$ with respect to
the one in $AdS_3$ example in \cite{kumar}.
An explicit solution for equations (\ref{AB-condition22}) and
(\ref{bianchi22}) is given as:
\be
A_{\psi} = 2\; {\rm{cos}}^2\theta\; {\rm{cos}}^2\phi,\;\;
A_{\omega} =  2 {\rm{cos}}^2\theta\; {\rm{sin}}^2\phi,
\label{Aa22}
\ee
and
\be
A_{\theta \psi} = - {\rm{sin}}2\theta\; {\rm{cos}}^2\phi,\;\;
A_{\theta\omega} = - {\rm{sin}}2\theta\; {\rm{sin}}^2\phi,\;\;
- A_{\phi \psi} = A_{\phi\omega} = {\rm{cos}}^2\theta\; {\rm{sin}}2\phi,\;\;
\label{Aab22}
\ee
for the choice of $S^5$ metric:
\be
   d\Omega_5^2 = d\theta^2 + {\rm{cos}}^2\theta\; d\phi^2 +
                 {\rm{cos}}^2\theta\; {\rm{cos}}^2\phi\; d\psi^2 +
        {\rm{cos}}^2\theta\; {\rm{sin}}^2\phi\; d\omega^2 +
                 {\rm{sin}}^2\theta\; d\gamma^2.
\label{deomega522}
\ee

Using the above expressions for $A_a$'s and $A_{ab}$'s,
the wave profile $H$ in the present example can be shown to satisfy:
\be
- {H_{, uu}\over 2} + {3\over 2} {H_{, u}\over u}
    - {1\over 2} H_{,i}^i - {1\over 2u^2\sqrt{g}}\partial_a
       (\sqrt{g}\partial^a H) = {4\mu^2 q^2\over u^4},
\label{H-equation22}
\ee
with a solution (independent of $x^i$'s) given by
\be
 H = - {\mu^2 q^2\over u^2} cos^2\theta.
\label{H22}
\ee
Once again, solution given in equations (\ref{Aa22}),
(\ref{Aab22}) and (\ref{H22}) can be
generalized further by multiplying $A$'s and $H$ with functions
$F(x^+)$ and $G(x^+)$ respectively, satisfying $G = F^2$.
We have therefore again presented an
explicit example of a   
gravitational wave in $AdS_5\times S^5$ background. 

We now show the connection of our
explicit solution given in equations (\ref{Aa22}), (\ref{Aab22})
and (\ref{H22}) with a a supersymmetric
D3-brane solution~\cite{biswas}.
To show this connection, we write
down the relevant $D3$ brane solution in a pp-wave background
(which is obtained by taking a `Penrose limit'\cite{penrose} on an
$AdS_3\times S^3\times R^4$ background of string theory). 
the D-brane solution is:
\bea
ds^2&=&f^{-{1\over2}}(2 dx^+dx^- -\mu^2{\sum_{i=1}^{4}}x_i^2(dx^+)^2+
(dx_5)^2+(dx_6)^2)\cr
& \cr
&+& f^{1\over2}{\sum_{a=1..4,7,8}}(dx_a)^2  \cr
& \cr
H_{+12}&=&H_{+34}= 2\mu, \cr
& \cr
F_{+ - 5 6 a}&=& \partial_a {f}^{-1},~~~~e^{2\phi} = 1,
\label{d3-bkgrd}
\eea
with $f$ being the Green function in six dimensional transverse
space with coordinates $x^{1,..,4}, x^7,x^8$. We write this Green 
function as: $f = (1 + {q^2\over r^4})$. We also mention that 
among the above coordinates, 
$x^{1..4}$ are also the pp-wave directions and 
$x^+, x^-$, $x^5, x^6$ are longitudinal directions
of the brane. 

To arrive at the metric and NS-NS 3-form of equations
(\ref{metric}) and (\ref{nsns3-form22}) (with $A_a$'s, $A_{ab}$'s
and $H$ as in equations (\ref{Aa22}), (\ref{Aab22}) and
(\ref{H22})) from the $D3$ brane solution in (\ref{d3-bkgrd}), we
now make the following coordinate transformations from the
Cartesian to radial and angular variables: 
\bea 
&&x_1 = r\; {\rm{cos}}\theta\; {\rm{cos}}\phi\; {\rm{cos}}\psi,\;\;
x_2 = r\; {\rm{cos}}\theta\; {\rm{cos}}\phi\; {\rm{sin}}\psi,\; \cr & \cr 
&&x_3 = r\; {\rm{cos}}\theta\; {\rm{sin}}\phi\; {\rm{cos}}\omega,\;\; 
x_4 = r\; {\rm{cos}}\theta\; {\rm{sin}}\phi\; {\rm{sin}}\omega, \cr &\cr 
&&x_7 = r\; {\rm{sin}}\theta\; {\rm{cos}}\gamma,\;\; 
x_8 = r\; {\rm{sin}}\theta\; {\rm{sin}}\gamma. 
\label{coordinates22} 
\eea 
Then taking the scaling
limit $r\rightarrow 0$, and defining $r = {q \over u}$,  one gets
the result given in equations (\ref{Aa22}), (\ref{Aab22}) and
(\ref{H22}).

We have therefore shown the connection of our gravitational wave
solution in this subsection with a $D3$ brane in a pp-wave background.
The background pp-wave itself follows from a Penrose limit on 
$AdS_3\times S^3\times R^4$ geometry with an appropriate 
3-form NS-NS or R-R flux. Just as in section-(2.1), we can also 
obtain other solutions for the wave-profile $H$ in equation 
(\ref{H-equation22}). It will again be of interest to find out 
which D-branes lead to a gravitational wave of this type in a 
singular limit $r\rightarrow 0$. 

Due to arguments similar to the ones 
in section-(2.1), namely that the
supersymmetry projections are independent of the Green function $f$, 
we expect our gravitational wave solution to be supersymmetric. 
In particular, since the $D3$ brane solution of equation (\ref{d3-bkgrd})
preserves $1/8$ supersymmetry, as discussed in section-(3.1)
of \cite{biswas}, we expect the gravitational
wave solution to preserve at least this much supersymmetry as well. 
Possible enhancement of supersymmetry can be analyzed by 
solving the Killing spinor equations. However, at this point we move
from Type IIB supergravity to study the 
gravitational waves in elven dimensional M-theory.

\section{$Ad_7\times S^4$ Gravitational Wave Solution in M-theory}

\subsection{$AdS_{7}$ Wave}

We now write down the gravitational wave solution in $AdS_7\times S^4$
backgrounds. These background configurations appear in M-theory
in a near horizon geometry of M5 brane solutions solving the
eleven-dimensional supergravity equations of motion.
The metric ansatz  for the gravitational wave solution that
we obtain has the following form:
\be
ds^2 = 4 q  \left\{ {du^2\over u^2} + {1\over u^2} (2 dx^+ dx^-
       + H(u, x^+, x^i, x^a) {dx^+}^2 + \sum_{i=1}^4 {dx^i}^2) +
       {1\over 4} d\Omega_4^2 \right\},
\label{metric3}
\ee
For $H=0$ this metric reduces to that of $AdS_7\times S^4$.
This theory, in addition, contains a 3-form field (or the corresponding
4-form flux). Note that we parameterize the space-time using coordinates
analogous to the
earlier section. In our case the 4-form flux has the following form:
\bea
F^{(4)} =  {16 \mu q^{3\over 2} \over u^3} A_{ij} (u, x^+, x^i, x^a)
            dx^+ \wedge du \wedge dx^i \wedge dx^j
          + {8\mu q^{3\over 2}\over u^2} A_{ija}dx^+\wedge dx^i
            \wedge dx^j \wedge dx^a \cr
& \cr
            - 3 q^{3\over 2}\sqrt{g} d\theta^1 \wedge d\theta^2
             \wedge d\theta^3 \wedge d\theta^4.\;\;\;\;\;\;\;\;\;
\label{4-form3}
\eea

Nonzero Ricci tensor components along $AdS_7$ directions, 
for the above metric, has a form:
\bea
   R_{++} = - {H_{, uu}\over 2} + {5\over 2}{H_{, u}\over u}
            - {6H\over u^2} - {1\over 2} H_{,i}^i
            - {2\over u^2 \sqrt{g}}\partial_a(\sqrt{g}
              \partial^a H),\cr
& \cr
   R_{+-} = R_{uu} = - {6\over u^2}, \;\;
   R_{ij} = - {6\delta_{ij}\over u^2}\; .
\label{ricci-ads7}
\eea
Ricci tensor components along $S^4$ directions are:
\be
   R_{a b} = 3 g_{a b}.
\label{ricci-s4}
\ee

We now derive constraints on $A_{ij}$, $A_{ija}$ and $H$ from the
eleven-dimensional supergravity equations of motion ($G_{\mu \nu}$
represents the eleven dimensional metric, as opposed to $g_{ab}$ which
denotes the metric on the $S_{4}$):
\be
   R_{\mu\nu} - {1\over 2} G_{\mu\nu} R =
        {1\over 12}\left(F_{\mu\alpha\beta\gamma}
       F_{\nu}^{\alpha\beta\gamma} - {1\over 8}
       G_{\mu\nu}F_{\alpha\beta\gamma\delta}
       F^{\alpha\beta\gamma\delta}\right),
\label{eom-g11}
\ee
and
\be
\partial_{\mu}(\sqrt{-G}F^{\mu\nu\rho\sigma}) + {1\over 1152}
 \epsilon^{\nu\rho\sigma\alpha_1\alpha_2\alpha_3\alpha_4
  \beta_1\beta_2\beta_3\beta_4}F_{\alpha_1\alpha_2\alpha_3\alpha_4}
  F_{\beta_1\beta_2\beta_3\beta_4} = 0,
\label{eom-4form}
\ee
where $F_{\mu\nu\rho\sigma}$, as mentioned earlier,
is a 4-form field strength in
the eleven-dimensional supergravity theory.

By using our ansatz for the metric and $F_{\mu\nu\rho\sigma}$
in equations (\ref{metric3}) and (\ref{4-form3}) we have,
for example:
\be
 R_{+-} - {1\over 2}g_{+-}R = - {1\over 96}g_{+-}F_{\mu\nu\rho\sigma}
                            F^{\mu\nu\rho\sigma} = -{9\over u^2}.
\ee
For all other components of $R_{\mu\nu}$, except $R_{++}$,
one gets identical field equations. For
the $(++)$ component, on the other hand, we get:
\be
    - {H_{, uu}\over 2} + {5\over 2} {H_{, u}\over u}
    - {1\over 2} H_{,i}^i - {2\over u^2\sqrt{g}}\partial_a
       (\sqrt{g}\partial^a H)
         =  {\mu^2} \left( \sum_a A_{ij}^2 +
            \sum_{ija} (A_{ija}A_{ij}^a    \right).
\label{h-condition3} 
\ee 
The 4-form field equation
(\ref{eom-4form}) implies the following conditions for quantities
$A_{ij}$ and  $A_{ija}$: \be
 \partial_j A_{ij} = 0,\;\; \partial_j A_{ija} = 0,
\label{A-condition31}
\ee
\be
 -{1\over \sqrt{g}}\partial_a(\sqrt{g}A_{ij}^a) + A_{ij}
  + {3\over 2}\epsilon^{ijkl}A_{kl} = 0.
\label{A-condition32} 
\ee 
From the Bianchi identity, one also has:
\be
 A_{i j a} = - \partial_{a} A_{{i j}}.
\label{bianchi3} 
\ee  
We now find that for the $S^4$ metric: \be
  d\Omega_4^2 = d\theta^2 + {\rm{sin}}^2\theta\;
                [d\phi^2 + {\rm{cos}}^2\phi\; d\psi^2
                + {\rm{sin}}^2\phi\; d\omega^2],
\label{deomega4}
\ee
$A_{i j}$ and $A_{i j a}$:
\be
A_{2 3} = A_{1 4} = - {\rm{cos}}\theta,\;\; A_{2 3 \theta} =
        A_{1 4 \theta} = - {\rm{sin}}\theta,
\label{Asolution-3}
\ee
and
\be
   H = - \mu^2 (\sum_{i=1}^4 x_i^2),
\label{Hsolution-3}
\ee
provide an explicit solution for equations:
(\ref{h-condition3}), (\ref{A-condition31}),
(\ref{A-condition32}) and (\ref{bianchi3}). 
Other examples can also be obtained, as in previous sections, by
multiplying $A$'s and $H$ by functions of $x^+$, as well as 
by solving equation (\ref{h-condition3}) directly.

We now show that the gravitational wave solution, with its explicit 
form as in equations (\ref{Asolution-3}) and (\ref{Hsolution-3}),
are obtained from a supersymmetric $M5$ brane solution in a pp-wave background
\cite{singh,alishahiha} by applying a limit, taking
$r\rightarrow 0$. The $M5$ brane solution is given as:
\bea
ds^2 = f^{-{1\over3}}(2 dx^+dx^- -\mu^2{\sum_{i=2,3,4,11}^{4}}
\tilde{x}_i^2(dx^+)^2+
\sum_{i=2,3,4,11}(dx_i)^2)
+ f^{2\over3}(dr^2 + r^2 d\Omega_4^2),  \cr
&\cr
F^{(4)} = 2\mu dx^+ \wedge ( d\tilde{x}_1\wedge d\tilde{x}_3
          \wedge d\tilde{x}_4 +
          d\tilde{x}_1 \wedge d\tilde{x}_2 
          \wedge d\tilde{x}_{11} ) +
          \epsilon_{m n l p q} \partial_q f d\tilde{x}^m\wedge 
           d\tilde{x}^n\wedge
          d\tilde{x}^l \wedge d\tilde{x}^p,\cr
&\cr
    f = (1 + {q^{3\over 2}\over r^3}),\;\;\;\;\;\;\;\;\;\;
\;\;\;\;\;\;\;\;\;\;\;\;\;\;\;\;\;\;\;\;\;\;\;\;\;\;\;\;\;\;\;
\label{M5-soln.} 
\eea 
where we have kept the longitudinal
coordinate indices in a manner such that its origin as a
dimensionally `oxidized' $D4$ brane~\cite{alishahiha} becomes more
clear. In our notation here, the five transverse coordinates,
denoted by $\tilde{x}^m$, 
representing $r$ and $d\Omega_4$, are given
as: $\tilde{x}^1, y^{1,..,4}$. Now, to obtain the gravitational wave
solution in equations (\ref{metric3}), (\ref{4-form3}),
(\ref{Asolution-3}), (\ref{Hsolution-3}), we define coordinates
as: 
\bea 
&&\tilde{x}_1 = r {\rm{cos}}\theta,\;\; 
y_1 = r {\rm{sin}}\theta\; {\rm{cos}}\phi\; {\rm{cos}}\psi,\;\; 
y_2 = r {\rm{sin}}\theta\; {\rm{cos}}\phi\; {\rm{sin}}\psi,\cr &\cr 
&&y_3 = r {\rm{sin}}\theta\; {\rm{sin}}\phi\; {\rm{cos}}\omega,\;\; 
y_4 = r {\rm{sin}}\theta\; {\rm{sin}}\phi\; {\rm{sin}}\omega, 
\label{coordinates3} 
\eea 
and apply the limit on
the transverse radius: $r\rightarrow 0$ while redefining: $ r =
4{q^{3\over 2}\over u^2}$. Identifications for the longitudinal 
coordinates used in (\ref{Asolution-3}) and (\ref{Hsolution-3}) are:
\be
x_1 \equiv \tilde{x}_2,\;\;x_2 \equiv \tilde{x}_3,\;\;
x_3 \equiv \tilde{x}_4,\;\;x_4 \equiv \tilde{x}_{11},\;\;
\label{ident-long}
\ee

Supersymmetry of the gravitational wave is once again
expected, due to the above connection with a $3/16$ supersymmetric
$M5$ brane. The supersymmetry of both $M5$\cite{singh} and 
the related $D4$\cite{alishahiha} branes have been obtained explicitly.
The projection conditions again turn out to be independent of $f$,
implying once more that the final solution is supersymmetric as well.

\subsection{Wave Profiles dependent on the $S^{4}$}

We now present another class of $AdS_7\times S^4$ gravitational wave
solution. The new solutions are obtained by
considering variations of the
flux ansatz~(\ref{4-form3}), allowing for the wave profile to be
dependent on the transverse $S^{4}$. To this end, consider the
following ansatz:
\bea 
F^{(4)} =  {32 \mu q^{3} \over u^5} B_{ai}
(u, x^+, x^i, x^a)
            dx^+ \wedge du \wedge dx^a \wedge dx^i
          + {8\mu q^{3}\over u^4} B_{abi}dx^+\wedge dx^a
            \wedge dx^b \wedge dx^i \cr
& \cr
            - 3 q^{3\over 2}\sqrt{g} d\theta^1 \wedge d\theta^2
             \wedge d\theta^3 \wedge d\theta^4,\;\;\;
\label{4-form4} 
\eea 
with $i$ running over $1,..,4$ and $x^a$'s are the four angular
coordinates. 
As before, we can derive constraints on the
undetermined functions. From the Bianchi identity
\be  
B_{[ai,b]}=
B_{abi}, 
\ee 
while the equation of motion $d\ast F = -F \wedge F$
yields
\be
 \partial_j B_{ja} = 0,\;\; \partial_j B_{abj} = 0,
\label{B-condition31} 
\ee 
\be
 B^{a}_{\phantom{i}i} =
 \frac{1}{6\sqrt{g}}\partial_{b}(\sqrt{g}B^{ba}_{\phantom{ii}i}).
\label{B-condition32} 
\ee 
Finally, we note that the $R_{++}$
equation of motion implies that
 \be
    - {H_{, uu}\over 2} + {5\over 2} {H_{, u}\over u}
    - {1\over 2} H_{,i}^i - {2\over u^2\sqrt{g}}\partial_a
       (\sqrt{g}\partial^a H)
         = \frac{{\mu^2 q^3}}{u^6} \left( 32 \sum_{ai} B_{ai}B^{a}_{i} +
            4\sum_{abi} (B_{abi}B^{ab}_{\phantom{ij}i})    \right).
\label{h-condition3b} 
\ee  
Writing the $S^{4}$ in coordinates such
that the metric is once again of the form (\ref{deomega4}), one
has the solution:
\be 
H = -\frac{4q^{3}\mu^{2}}{u^{4}}{\rm{sin}}^{2}\theta, 
\ee 
with flux determined by 
\begin{eqnarray} 
B_{\psi 1} =
- {\rm{sin}}^{2}\theta\;{\rm{cos}}^2\phi, &  B_{\omega 1}
= - {\rm{sin}}^{2}\theta\; {\rm{sin}}^2 \phi,
\end{eqnarray} 
and 
\begin{eqnarray}
B_{\theta \psi 1} & = & {\rm{sin}}2\theta\; {\rm{cos}}^{2} \phi, \nonumber \\
B_{\phi \psi 1} & = & - {\rm{sin}}^{2}\theta\; {\rm{sin}}2\phi, \\
B_{\theta \omega 1} & = & {\rm{sin}}2\theta\; {\rm{sin}}^2\phi, \nonumber \\
B_{\phi \omega 1} & = & {\rm{sin}}^{2}\theta\; {\rm{sin}} 2\phi. \nonumber
\end{eqnarray}
This gravitational wave can also be found as the near-horizon
limit of a supergravity solution describing a stack of M5 branes
in the plane wave background with 20 supersymmetries \cite{mas}.
The resulting configuration has eight supersymmetries and is given
by 
\bea 
ds^2 = f^{-{1\over3}}(2 dx^+dx^-
-\frac{\mu^2}{4}{\sum_{m=5}^{8}}x_m^2(dx^+)^2+ \sum_{i=1}^{4}(dx_i)^2) +
f^{2\over3}\sum_{m=5}^{9}(dx^{m})^2, \cr &\cr F^{(4)} = \mu dx^+
\wedge ( dx^5\wedge dx^6\wedge dx^1 +
          dx^7 \wedge dx^8 \wedge dx^{1} ) +
          {\epsilon_{m n l p q} \over 4!} \partial_q f dx^m\wedge dx^n\wedge
      dx^l \wedge dx^p,\cr
&\cr
    f = (1 + {q^{3\over 2}\over r^3}).\;\;\;\;\;\;\;\;\;\;
\;\;\;\;\;\;\;\;\;\;\;\;\;\;\;\;\;\;\;\;\;\;\;\;\;\;\;\;\;\;\;
\label{M5-soln2.} 
\eea 
Upon parameterizing the transverse coordinates
($x^5,...,x^8, x^9$ $\equiv$ $y^1,..,y^4,\tilde{x}_1$) 
exactly as above  in equation (\ref{coordinates3}), defining
the coordinate $u$ as before, and taking the appropriate near
horizon limit, one recovers the gravitational wave solution
outlined above. 

Supersymmetry of the solution can be discussed along the lines of 
other examples in section-2. We now go over to the gravitational wave
solution in $AdS_4\times S^7$ background. 

\section{$AdS_4\times S^7$ Solution}

Gravitational waves in $AdS_4$ backgrounds are of particular interest,
due to their connection with the physics in four dimensions. In
pure gravity theory, the gravitational waves in such backgrounds
require the presence of cosmological constant term. In eleven dimensional
M-theory that we are considering, one does not have any such 
cosmological constant term and the background $AdS_4$ 
is accompanied by an $S^7$ in order to compensate for the opposite
Ricci curvature terms. Phenomenological consequences of such a 
gravitational wave in $AdS_4\times S^7$ background will be 
also of interest to examine along the lines of \cite{carmelli}.

We now give an example of a  gravitational wave in $AdS_4\times S^7$
background. Later on, in this section, we also show the connection of our
solution with certain supersymmetric `localized' M2 branes
of \cite{mas} in the same way as was done above for other branes.
The metric is now written as:
\be
ds^2 =  {q\over 4}  \left\{ {du^2\over u^2} + {1\over u^2} (2 dx^+ dx^-
       + H(u, x^+, x, x^a) {dx^+}^2 + {dx}^2) +
       4 d\Omega_7^2 \right\},
\label{metric4}
\ee
The 4-form flux is of the form:
\be
F^{(4)} = {3 q^{3\over 2}\over 8 u^4} dx^+ \wedge dx^- \wedge dx
          \wedge du - {\mu q^{9\over 4}\over 4\sqrt{2}}
          {A_{ab}\over u^{5\over 2}}dx^+ \wedge du \wedge dx^a
          \wedge dx^b + {\mu q^{9\over 4}\over 2\sqrt{2}}
           {A_{a b c}\over u^{3\over 2}} dx^+ \wedge dx^a
           \wedge dx^b \wedge dx^c .
\label{4-form42}
\ee
where powers of $q$ are chosen appropriately to have q-independent
solution for $A_{a b}$'s etc. below. For the Ricci curvature 
components we now have:
\bea
   &&R_{++} = - {H_{, uu}\over 2} + {H_{, u}\over u}
            - {3H\over u^2} - {1\over 2} H_{,i}^i
            - {1\over 8u^2 \sqrt{g}}\partial_a(\sqrt{g}
              \partial^a H), \cr
& \cr
   &&R_{+-} = R_{uu} = R_{xx}= - {3\over u^2},\;\;
   R_{ab} = 6 \delta_{a b}.
\label{ricci-ads4}
\eea
with $a,b$ denoting the $S^7$ coordinates.
Equations of motion then simplify to:
\be
 {1\over \sqrt{g}}\partial_c(\sqrt{g}A^{abc}) + 5 A^{ab} = 0,\;\;
 {1\over \sqrt{g}}\partial_b(\sqrt{g}A^{ab}) = 0.
\label{A-condition4}
\ee
and
 \be
    - {H_{, uu}\over 2} +  {H_{, u}\over u}
    - {1\over 2} H_{,i}^i - {1\over 8u^2\sqrt{g}}\partial_a
       (\sqrt{g}\partial^a H)
         =  {1\over 32} {\mu^2 q^{3\over 2}\over u^3}
            \left( \sum_{a, b} A_{ab} A^{ab} +
            {1\over 3}\sum_{a, b, c} (A_{abc}A^{abc}    \right).
\label{h-condition4}
\ee
Bianchi identity gives:
\be
 A_{a b c} =  \partial_{[a} A_{b c]}.
\label{bianchi4}
\ee

One can also write an explicit solution for all the conditions, namely
equations  (\ref{A-condition4}), (\ref{h-condition4}), and
(\ref{bianchi4}). For this we write down a metric on $S^7$ as:
\be
    ds^2 = d\theta^2 + {\rm{cos}}^2\theta\; [d\phi^2 +
        {\rm{cos}}^2\phi\; d\psi^2 + {\rm{sin}}^2\phi\; d\omega^2]
    + {\rm{sin}}^2\theta\; [d\gamma^2 + {\rm{sin}}^2\gamma\; d\eta^2
        + {\rm{sin}}^2\gamma\; {\rm{sin}}^2\eta\; d\beta^2 ].
\label{domega7}
\ee
Our solution for $A_{ab}$ and $A_{abc}$ are then:

\begin{eqnarray}
&& A_{\psi \theta} = \rm{cos}\theta\; \rm{cos}^2\phi\; \rm{cos}\gamma,
 \qquad \qquad
 A_{\omega \theta} = \rm{cos}\theta\; \rm{sin}^2\phi\; \rm{cos}\gamma,
 \cr &\cr
 &&A_{\psi \gamma} = - \rm{cos}^2\theta\; \rm{cos}^2\phi\;
                            \rm{sin}\theta\; \rm{sin}\gamma, \qquad
 \quad \!
 A_{\omega \gamma} = - \rm{cos}^2\theta\; \rm{sin}^2\phi\;
                   \rm{sin}\theta\; \rm{sin}\gamma, \cr &\cr
 &&A_{\phi \psi} = - \rm{cos}^2\theta\; \rm{sin}\theta\;
                  \rm{sin}\phi\; \rm{cos}\phi\; \rm{cos}\gamma, \quad
 A_{\phi \omega} = \rm{cos}^2\theta\; \rm{sin}\theta\; \rm{sin}\phi\;
                  \rm{cos}\phi\; \rm{cos}\gamma,
\label{Aab4}
\end{eqnarray}
and
\bea
&&A_{\theta \psi \gamma} = \rm{sin}^2\theta\; \rm{cos}\theta\;
                       \rm{cos}^2\phi\; \rm{sin}\gamma, \qquad
A_{\theta \omega \gamma} = \rm{sin}^2\theta\; \rm{cos}\theta\;
                       \rm{sin}^2\phi\; \rm{sin}\gamma,\;\;\cr
&\cr
&&A_{\phi \psi \theta} = - \rm{cos}^3\theta\; \rm{sin}\phi\;
                       \rm{cos}\phi\; \rm{cos}\gamma, \quad \;
A_{\phi \psi \gamma} = \rm{cos}^2\theta\; \rm{sin}\theta\; \rm{sin}\phi\;
                      \rm{cos}\phi\; \rm{sin}\gamma,\;\;\cr
&\cr
&&A_{\phi \omega \theta} = \rm{cos}^3\theta\; \rm{sin}\phi\;
                         \rm{cos}\phi\; \rm{cos}\gamma, \qquad
A_{\phi \omega \gamma} = - \rm{cos}^2\theta\; \rm{sin}\theta\; 
                        \rm{sin}\phi\;
                       \rm{cos}\phi\; \rm{sin}\gamma.\;\;
\label{Aabc4}
\eea
For the wave profile ($H$) we have:
\be
  H = - {\mu^2 q^{3\over 2}\over 8 u} cos^2\theta .
\label{H4}
\ee

We now show that the $AdS_4\times S^7$ gravitational wave solution,
characterized by functions $A_{a b}$, $A_{a b c}$ and $H$
in equations (\ref{Aab4}), (\ref{Aabc4}) and (\ref{H4}), is obtained from
an $M2$ brane solution~\cite{mas} in a pp-wave background in a near horizon 
geometry. The $M2$ brane solution is given as:
\begin{eqnarray}
&&ds^2 = f^{-{2\over3}}(2 dx^+dx^- - H (dx^+)^2+ (dx)^2)
+ f^{1\over3}(\sum_{i=1}^8 dx_i^2),  \cr
& \cr
&&F^{(4)} =  dx^+ \wedge ( \mu_1 dx_1\wedge dx_2+
          \mu_2 dx_3 \wedge dx_4 + \mu_3 dx_5 \wedge dx_6 ) \wedge dx_8,  \cr
&\cr
&&    f = (1 + {q^{3}\over r^6}),\;\;\;\;\;\;\;\;\;\;\;\;\;\;\;\;
\;\;\;\;\;\;\;\;\;\;\;\;\;\;\;\;\;\;\;\;\;
\label{M2-soln.}
\end{eqnarray}
with $H$ in equations (\ref{M2-soln.}) being: 
\be
  H = - {\mu_1^2\over 4} (x_1^2 + x_2^2)
      - {\mu_2^2\over 4} (x_3^2 + x_4^2)
      - {\mu_3^2\over 4} (x_5^2 + x_6^2). 
\label{HM44}
\ee

The general solution above, for the $M2$ brane,
has three independent parameters $\mu_1$,
$\mu_2$ and $\mu_3$. However, to obtain the gravitational wave solution 
above, in equations (\ref{Aab4}), (\ref{Aabc4}) and (\ref{H4}) we have set
the parameter $\mu_3$ to zero. Moreover, we have also set $\mu_1 = \mu_2$.
Other supersymmetric solution $\mu_1 = -\mu_2$ in \cite{mas}
is similar to the one we have written in
(\ref{Aab4}), (\ref{Aabc4}) and (\ref{H4}), only change being that 
$\omega$ components in $A_{ab}$ and $A_{abc}$ are changed by a minus sign.

Now, to obtain the gravitational wave solution obtained above,
we make the following coordinate transformations:
\bea
&&x_1 = r\; {\rm{cos}}\theta\; {\rm{cos}}\phi\; {\rm{cos}}\psi,\;\;
x_2 = r\; {\rm{cos}}\theta\; {\rm{cos}}\phi\; {\rm{sin}}\psi,\;\cr
&\cr
&&x_3 = r\; {\rm{cos}}\theta\; {\rm{sin}}\phi\; {\rm{cos}}\omega,\;\;
x_4 = r\; {\rm{cos}}\theta\; {\rm{sin}}\phi\; {\rm{sin}}\omega, \;\cr
&\cr
&&x_5 = r\; {\rm{sin}}\theta\; {\rm{sin}}\gamma\; {\rm{cos}}\eta,\;\;
x_6 = r\; {\rm{sin}}\theta\; {\rm{sin}}\gamma\; {\rm{sin}}\eta\; 
{\rm{cos}}\beta, \cr
&\cr
&&x_7 = r\; {\rm{sin}}\theta\; {\rm{sin}}\gamma\; {\rm{sin}}\eta\; 
{\rm{sin}}\beta,\;\;
x_8 = r\; {\rm{sin}} \theta\; {\rm{cos}}\gamma,
\label{coordinates4}
\eea
and take the limit $r\rightarrow 0$ while also defining
$r = {q^{3\over 4}\over \sqrt{2}} u^{-{1\over 2}}$.
We then obtain the solution in equations (\ref{Aab4}),
(\ref{Aabc4}) and (\ref{H4}). The supersymmetry of the gravitational 
wave solution is once again expected , following similar arguments 
as in previous sections. 

\section{Conclusion}

In this paper we have obtained several examples of gravitational
wave solutions in string theory and M-theory. A new feature of our 
solution is the presence of new p-form fluxes that are present. 
These fluxes also dictate the form of the wave profiles that one
obtains by solving the wave equations. We have also presented 
many examples.  The general structure of these examples 
have been dictated by certain D-branes
and M2, M5 branes in pp-wave backgrounds. It should certainly be
possible to extend these solutions further and obtain the
gravitational waves in $AdS$ backgrounds in large number of other
possible cases, coming from various other brane solutions. In this 
context, D-branes of maximally supersymmetric pp-wave backgrounds
will be particularly interesting to study, as the corresponding 
gravitational wave may provide interpolation between 
$AdS_5\times S^5$ geometry and a pp-wave with maximal supersymmetry,
by using procedures outlined in previous sections. 
Such an interpolating solution may have an interesting
interpretation in $N=4$ supersymmetric gauge theories as well.
Further, it would also be interesting to compute the Penrose limit
along null geodesics on the spheres of the space-times given above.
Such an analysis was carried out earlier in the case of Kaigorodov
space-times~\cite{patri}. which can be derived from our ansatz above for the
$AdS_{4} \times S^{7}$ wave by setting $\mu = 0 $. One should in
principle find a supersymmetric plane wave, but the form of the
resulting wave profile would be highly non-trivial. It should also be
noted that a Penrose limit along the perturbed $AdS$ part of the
metrics above should also be non-trivial, since the space-times are not
conformally flat. 

Our results can also possibly be of use for discussing holography in a
more general context than pure $AdS_m\times S^n$  type
solutions~\cite{brecher}. An interesting example from this point
of view may be the one obtained from a $D3$ brane, with wave
profile given in equation (\ref{H1'}). We notice that the boundary
geometry in this case is a four dimensional pp-wave. What
implications this observation may have on the CFT structure is
worth examining. In particular, it may be interesting to find out
the meaning of an additional parameter ($\mu$), in the examples
discussed in this paper, in the CFT side, same way as the radius
of curvature for $AdS_5\times S^5$ is related to the rank of the
gauge group $N$.

{\bf Note Added:} After the submission of this paper to the 
archive, we have also come across another paper \cite{kerimo}
where gravitational wave solutions have been obtained in 
AdS backgrounds.

{\bf Acknowledgment:} We would like to thank B. Chandrasekhar for
useful discussions. HKK is grateful to St John's College for a
Benefactor's Scholarship. 

%%%%%%%%%%%%%%%%%%%%%%%%%%%%%%%%%%%%%%%%%%%%%%%%%%%%%%%%%%%%%%%%%%%%%%%%
%                       REFERENCES                                     %
%%%%%%%%%%%%%%%%%%%%%%%%%%%%%%%%%%%%%%%%%%%%%%%%%%%%%%%%%%%%%%%%%%%%%%%%
%\newpage

\renewcommand{\thefootnote}{\arabic{footnote}}
\setcounter{footnote}{0}

%\tableofcontents
%%%%%%%%%%%%%%%%%%%%%%%%%%%%%%%%%%%%%%%%%%%%%%%%%%%%%%%%%%%%%%%%%%%%%%
%\section{Introduction}
%%%%%%%%%%%%%%%%%%%%%%%%%%%%%%%%%%%%%%%%%%%%%%%%%%%%%%%%%%%%%%%%%%%%%%

\end{document}